\documentclass[amsmath,amssymb,aps,floatfix,jcp,superscriptaddress,preprint,11pt]{revtex4-2}
\usepackage[utf8]{inputenc} 
\usepackage[T1]{fontenc}    
\usepackage{amsfonts}       
\usepackage{booktabs}       
\usepackage{hyperref}       
\usepackage{url}            
\usepackage{microtype}      
\usepackage{nicefrac}       
\usepackage{paralist}       
\usepackage{graphicx}
\usepackage{float}
\usepackage{wrapfig}
\usepackage[dvipsnames]{xcolor}
\usepackage{ulem}
\usepackage{tikz}
\usetikzlibrary{arrows,shapes,angles,quotes}
\usetikzlibrary{decorations.pathreplacing}
\usepackage{fullpage}

\usepackage{lmodern}
\usepackage{ulem}

\usepackage{subcaption}

\begin{document}

\title{Feature Selection for High-Dimensional Neural Network Potentials with the Adaptive Group Lasso}

\author{Johannes Sandberg$^{1,2,3}$, Thomas Voigtmann$^{1,2}$, Emilie Devijver$^4$, Noel Jakse$^3$}
\address{$^1$Institut für Materialphysik im Weltraum, Deutsches Zentrum für Luft- und Raumfahrt (DLR), 51170 	
  Köln, Germany }
\address{$^2$Department of Physics, Heinrich-Heine-Universität Düsseldorf, Universitätsstraße 1, 40225 Düsseldorf, Germany}
\address{$^3$Université Grenoble Alpes, CNRS, Grenoble INP, SIMaP
  F-38000 Grenoble, France}
\address{$^4$Univ. Grenoble Alpes, CNRS, Grenoble INP, LIG, F-38000 Grenoble, France}

\begin{abstract}
Neural network potentials are a powerful tool for atomistic simulations, allowing to accurately reproduce \textit{ab initio} potential energy surfaces with computational performance approaching classical force fields.
A central component of such potentials is the transformation of atomic positions into a set of atomic features in a most efficient and informative way.
  In this work, a feature selection method is introduced for high dimensional neural network potentials, based on the Adaptive Group Lasso (AGL) approach.
  It is shown that the use of an embedded method, taking into account the interplay between features and their action in the estimator, is necessary to optimize the number of features.
The method's efficiency is tested on three different monoatomic systems,
including Lennard-Jones as a simple test case, Aluminium as a system characterized by predominantly radial interactions, and Boron as representative of a system
with strongly directional interactions.
The AGL is compared with unsupervised filter methods and found to perform
consistently better in reducing the number of features needed to reproduce
the reference simulation data.
{In particular, our results show the importance of taking into
  account model predictions in feature selection for interatomic potentials.}
\end{abstract}

\maketitle


\section{Introduction}
\label{Sec:Introduction}
{During the last decade, Machine Learning Interaction Potentials (MLIPs) have become a commonplace method for Molecular Dynamics simulations in material science and chemistry \cite{behler-2016-perspective, behler-2022-review}, following a broader trend of data-driven approaches in material science \cite{Schmidt-2019-review, Choudhary-2022-review}.
\textit{Ab initio} simulations, using for instance Density Functional Theory (DFT) force calculations \cite{Hafner-2008}, have good accuracy and broad applicability, but suffer from poor scalability.
  Being trained to reproduce \textit{ab initio} forces and energies, MLIPs were shown to combine many of the benefits of \textit{ab initio} with the scalability and performance of classical force fields \cite{EAM,MEAM,ReaxFF}}, thereby opening up new avenues of research into nucleation \cite{Jakse-2023-Al,nucleation-H2O,Al-Cu-NNP-2020}, structure-property relationship in alloys \cite{Al-Mg-Si-NNP-2021,Al-Cu-Mg-Zn-NNP-2022}, and amorphous solids \cite{Li-Si-Amorphous-NNP, Li3PO4-Amorphous-NNP} to name a few.

{
  A wide variety of MLIPs have been proposed, often relying on a local decomposition of the high dimensional potential energy into a sum of local contributions.
  Methods such as the spectral neighbor analysis potential \cite{SNAP} rely on a linear regression over a set of nonlinear descriptors of the local atomic environment.
  Nonlinear dependencies can be added by the use of kernel regression, as in the Gaussian approximation potential \cite{GAP, Gaussian-Process-Regression}, or by using Neural Networks (NN) as in the deep potential framework \cite{DeepMD} and the high dimensional neural network potential \cite{behler-parrinello-2007}.
  More recently, methods based on graph neural networks have seen a lot of traction \cite{SchNET}, including methods based on equivariant transformations \cite{NequIP}.
  Attempts have also been made to go beyond local interaction in what has been referred to as the third and fourth generations of machine learned potentials \cite{4-gens}.
}
For most MLIPs, it is necessary to transform the bare atomic coordinates into a set of atomic descriptors \cite{behler-perspective-2016} describing the local environment of each atom.
The purpose of this transformation is to enable a local description, ensure invariance to local symmetry transformations, and to guarantee that the input to the Machine Learning (ML) model is of constant dimension, even as the number of atomic neighbors can change during a simulation.

Computing the descriptors is often the main time consuming part of applying a NN Potential (NNP), compared to the NN evaluation and backpropagation.
As such, care is needed when designing the set of atomic features, and in particular one has to weight the need for a detailed description of the atomic environment against the additional computational cost of having a large feature space.
There is also some evidence that larger feature sets can negatively impact generalization \cite{wACSF}.
{Feature selection \cite{FS-review-2014} allows for a data driven way of designing such feature sets by identifying those features out of a larger collection that are the most relevant, and discarding redundant ones.}
The simplest approach to feature selection are filter methods.
Such methods select features by looking only at the dataset, before training takes place, and are as such model independent.
Imbalzano \textit{et~al.} \cite{imbalzano} proposed three such methods for use with MLIPs.
Two of these are based on minimizing the Pearson Correlation (PC), and maximizing the Euclidean distance,  respectively between the selected features.
The third one is based on the CUR decomposition \cite{CUR}, which can be regarded as an analogue of the singular value decomposition, constructing a low-dimensional representation of the data matrix but using only rows (columns) of the original matrix chosen such that the reconstruction error is minimized.

Filter methods can be contrasted with embedded methods, wherein the feature selection process is integrated into the training of a specific model.
Such an embedded approach allows for explicitly taking into account model predictions, as well as interaction between different features \cite{LassoNET}.
A famous embedded method is the lasso \cite{LASSO}, based on regularization using the $L1$ norm of the input parameters of a linear model.
Lasso has previously been used to construct MLIPs for a variety of elements based on ridge regression \cite{Seko-LASSO, Seko-ElasticNet}, and has been applied beyond MLIPs to predict directly material properties starting from large sets of material descriptors \cite{Ghiringhelli-2017}.
The latter led to the development of the SISSO method \cite{SISSO} in the framework of materials discovery, where features are subjected to an initial screening based on their correlation to the target property, before being further selected using the lasso, allowing for selection from more than billions of candidate material descriptors.
However, as it induces sparsity at the level of individual parameters, lasso is not applicable as a feature selection method for NNPs.

While much of the focus for feature selection has traditionally been on linear regression, likely owing to the nonlinear nature of NNs, recent works have tried to extend methods to the nonlinear case.
Methods based on the Group Lasso (GL) has been applied to NNs as early as 2017 \cite{original-GL}.
It has, however, been shown that this direct application of GL to NNs cannot consistently discard truly irrelevant features, a problem that can be avoided by using an adaptive penalty for an Adaptive GL (AGL) approach \cite{AGL}.
Another recent method is LassoNet \cite{LassoNET}, adding bypass connections from each input variable to the NN output, applying a lasso penalty on the bypass weights and using them to constrain the maximum values of the input weights.
{This change in architecture, however, deviates from the simple networks used in most common NNP implementations, while also introducing an additional hyperparameter that in principle needs to be tuned.
  For these reasons the AGL might be more directly suitable for NNPs.
}

In this article, we introduce an approach of feature selection based on the AGL method applied to High Dimensional NNPs (HDNNPs), with the aim of showing that the use of a method that takes into account the interplay between features in the specific estimator allows for better selection of atomic fingerprints.
{This type of NNP model is known to work well for many systems, and has been well studied, making it a natural framework for our study.}
We consider three different systems: \textit{Lennard-Jones} (LJ), serving as a simple and well known generic model whose analytic expression has no explicit angular dependence; \textit{Aluminium} (Al), which serves as a relatively simple sp bonding metal; \textit{Boron} (B), which is known to have a particularly complex structure with a high degree of directional covalent bonding \cite{beta-B-review, B-review-2009}.
{We find that for Al the AGL method is competitive with filter methods.
  For the other systems it is explicitly shown by example how the filters can fail to select features that are necessary, while they are discovered by our method, illustrating the advantage of an embedded feature selection approach.}

The remainder of the article is as follows.
Section~\ref{sec:method} provides background on our datasets, the HDNNP approach, the AGL method, and the computational tools used.
Section~\ref{sec:results} covers the results of training HDNNPs with AGL, comparing to the CUR and PC methods, as well as simulations used to test the effect of the reduced feature sets in production.
Finally, section~\ref{sec:conclusion} provides the main conclusions and outlook of the paper.

\section{Method}
\label{sec:method}

\subsection{Datasets}
\label{sec:Datasets}

\begin{figure}
  \centering
  \includegraphics[width=0.45\textwidth]{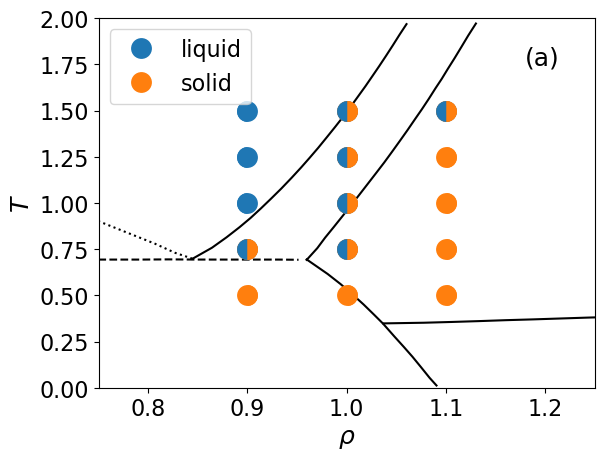}
  \includegraphics[width=0.45\textwidth]{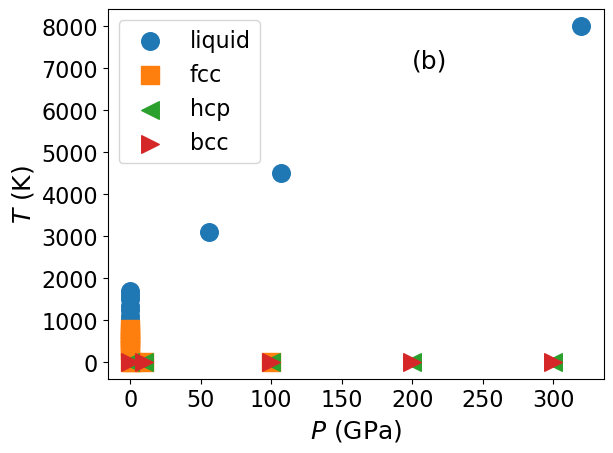}
  \includegraphics[width=0.45\textwidth]{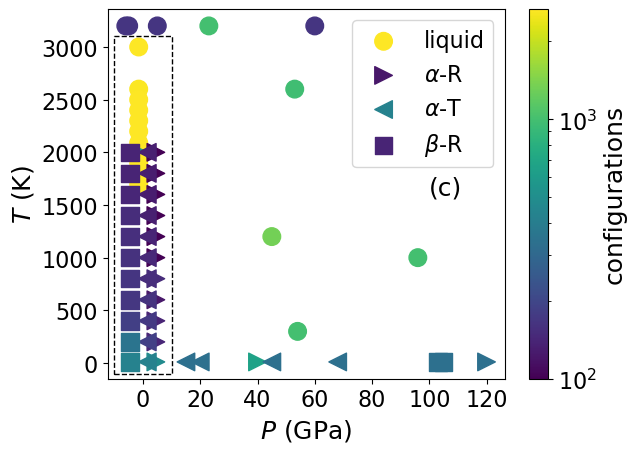}
  \caption{\label{fig:LJ-dataset}
    Thermodynamic points sampled in the construction of our datasets.
    Colors, symbols indicate respectively simulations started in different
    thermodynamic phases. (a) Temperature-density ($T$-$\rho$) phase diagram for LJ. For points with both colors, two
    separate simulation sets were included. (b) Temperature-pressure ($T$-$P$) phase diagram for Al. (c) Temperature-pressure ($T$-$P$) phase diagram for B. Points on the
    $P=0$ line, inside the dashed rectangle, have been shifted
    horizontally for readability. Here, colors represent the number of
    states sampled.}
\end{figure}

A first step of training a HDNNP is to construct a dataset of reference structures.
The dataset for LJ was extracted from a set of LAMMPS \cite{LAMMPS} simulations of $256$ atoms at temperatures ranging from $0.5$ to $1.5$ (LJ units), and densities $0.9$ to $1.1$, in both solid (fcc) and liquid configurations.
{We use the standard LJ pair potential, given for interatomic distance $r<r_c$ by
\begin{equation}
  V = 4\epsilon \left(\left(\frac{\sigma}{r}\right)^{12} - \left(\frac{\sigma}{r}\right)^6\right).
\end{equation}
All the simulations are performed with parameters $\sigma=\epsilon=1$, particle mass $m=1$, and cutoff radius $r_c=2.8$.
Figure~\ref{fig:LJ-dataset}(a) shows the thermodynamic states included in the dataset.
Each thermodynamic state was sampled $1333$ times, with an interval of $0.3$ time units ($300$ timesteps), for a total of $~28000$ configurations.
Note that the coexistence lines in figure~\ref{fig:LJ-dataset}, reproduced from \cite{LJ-phase-diagram}, are valid in the limit of infinite cutoff, and merely included as visual guide.
}

In the case of Al, our reference data is the same as in our previous article \cite{Jakse-2023-Al}.
This dataset consists of $24300$ configurations extracted from DFT-based \textit{Ab Initio} Molecular Dynamics (AIMD) simulations performed in VASP \cite{VASP} using the LDA functional \cite{LDA} in an augmented plane wave framework with a cutoff of $241$ eV.
Configurations in the dataset cover fcc, bcc, and hcp crystalline states, and the liquid, at a variety of temperatures and pressures the details of which we refer to the original article \cite{Jakse-2023-Al}.
{Figure~\ref{fig:LJ-dataset}(b) shows the thermodynamic points sampled to construct the dataset.
  Liquid states, and fcc crystals at ambient pressure were sampled $1000$ times each. The remaining crystal states were each sampled $100$ times.
}

For B, we extract reference configurations from the AIMD trajectories used in \cite{Jakse-2014-B}, complemented with additional simulations for $\alpha$-rhombohedral, $\alpha$-tetragonal, and $\beta$-rhombohedral crystals at temperatures ranging from $10$K to $2000$K in steps of $200$K, extracted from the \textit{Materials Project} database \cite{materials-project}.
Additional high-pressure simulations were also included, to probe the short-range interaction.
{Figure~\ref{fig:LJ-dataset}(c) shows the thermodynamic state of each simulation trajectory, with the number of configurations drawn from it.
Each trajectory was sampled with an interval of $45$ fs ($30$ timesteps), for a total of  $45000$ configurations.}
These simulations were performed using the Perdew Wang GGA functional \cite{Perdew-Wang-GGA} with a $300$ eV augmented plane wave cutoff sampling only the $\Gamma$ point, { for consistency with \cite{Jakse-2014-B}.}

{
In all cases, the simulations were performed in an NVT ensemble with a Nosé thermostat controlling the temperature, and pressure controlled by fixing the volume of the simulation box.
To ensure sampling of equilibrium states, each trajectory was preceded by an equilibration period ranging from 500 time units for LJ, and 100 to 200 ps for Al and B.
}

\subsection{HDNNPs}
The interaction between atoms in a material is frequently described in terms of a potential, depending in principle on the positions of all atoms in the many-particle system.
This interaction is often short-sighted, and can be treated as sum of atomic contributions depending only on the local structure of each atom, within an appropriate cutoff radius $r_c$
\begin{equation}
  E_\mathrm{total} = \sum_{i=1}^{N_\mathrm{atoms}} E_i.
\end{equation}
A HDNNP \cite{behler-parrinello-2007,behler-tutorial-2015} is constructed from this decomposition by assigning a NNP to each species of atom, mapping between the local environment and the corresponding atomic energy contribution $E_i$.
The input to the HDNNP are the atomic positions, which are transformed into a fingerprint vector for each atom, serving as input to the atomic NNP.
Training then consists of fitting the full HDNNP to the total potential energy obtained from \textit{ab initio}.
Often the derivative of the HDNNP is fitted to the \textit{ab initio} forces as well, but for simplicity in focusing on the feature selection and following our previous work \cite{Jakse-2023-Al}, we train only to the energies in this work.

There are many options in choosing atomic descriptors, with \cite{behler-perspective-2016} offering a brief overview of some common types.
In this work, we use the Behler-Parrinello symmetry functions (SF) \cite{behler-fingerprints-2011}, which is the conventional choice for HDNNPs.
These consist of the radial $G^2$ and angular $G^5$ SFs defined by
\begin{equation}
  \label{eq:G2}
  G^2_i = \sum_j e^{-\eta(R_{ij}-R_s)^2}f_c(R_{ij})
\end{equation}
\begin{equation}
  \label{eq:G5}
  G^5_i = 2^{1-\zeta}\sum_{j,k}(1 + \Lambda\cos\theta_{ijk})^\zeta
  e^{-\eta(R_{ij}^2+R_{ik}^2+R_{jk}^2)}f_c(R_{ij})f_c(R_{ik})f_c(R_{jk})\;.
\end{equation}
Here, $R_{ij}$ is the distance between atoms $i$ and $j$, $\theta_{ijk}$ is the angle between atoms $j$ and $k$ with respect to atom $i$, and $f_c(R_{ij})$ is defined as $0$ for $R_{ij} > r_c$ and for $R_{ij}<r_c$ as a polynomial going smoothly to 0 at the neighborhood cutoff $R_{ij} = r_c$.
The parameters $\eta$, $\zeta$, $\Lambda$, and $R_s$ allow for defining a set of {features} by assigning these parameters different values.
{
Here the initial featuresets are generated by selecting parameter values on a grid, akin to the procedures described in \cite{wACSF, imbalzano}, with the aim of being sensitive to a range of interatomic radii and angles.
The exact SF parameter values used can be found in the supplementary material \cite{si}.
}

\subsection{Feature Selection}
\label{sec:FS}
{
  The main hindrance in applying feature selection methods based on the L1 norm to NNs is the fact that the L1 norm acts on individual weights.
  In a NN, several weights are associated with each feature, and so to do feature selection we need to penalize these weights as a group.
The GL replaces the L1 norm with Euclidean norms over groups of parameters.}
As the Euclidean norm of a parameter group vanishes if and only if all those parameters vanish, this allows for selecting or discarding groups of parameters simultaneously.
To select features for NNs using GL we take the groups to be the input weights of feature $i$, $\omega_{i,[:]}^0$, with the corresponding Euclidean norm $|\omega_{i,[:]}^0|$. 
During training we then optimize the objective function
\begin{equation}
  \label{eq:GL}
  \mathrm{obj}(W) = L(W) + \frac{\lambda}{N}\sum_{i=1}^N |w^0_{i,[:]}|,
\end{equation}
with $L$ being some loss function{, in our case the Mean Square Error (MSE)}, $W$ being the weights of the neural network, $N$ being the number of inputs, and $\lambda$ being a regularization parameter used to tune the relative strength of the feature selection.
A challenge in performing this optimization is the fact that the second term in (\ref{eq:GL}), called the penalty, is non-smooth.
In \cite{Zhang-GL-2019} a smoothed approximation of (\ref{eq:GL}) is used, but here the non-smooth optimization problem is instead solved directly using a proximal gradient descent algorithm, following \cite{Feng-2019}.

The adaptive version of the algorithm \cite{AGL} uses a separate regularization parameter for each individual weight group.
This adapted penalty is constructed from an initial training run using the non-adaptive penalty.
The training is then redone with the new penalty, optimizing
\begin{equation}
  \label{eq:AGL}
  \mathrm{obj}(W) = L(W) + \frac{\lambda}{N}\sum_{i=1}^N \frac{|w^0_{i,[:]}|}{|\hat{w}^0_{i,[:]}|}
\end{equation}
with $\hat{w}^0_{i,[:]}$ being the values of $w^0_{i,[:]}$ obtained during the initial training run with the non-adaptive penalty.
Depending on the value of $\lambda$, some features will have their weights go to zero during training, and can thus be discarded.
This allows for selecting features by performing a search over this single parameter.

\subsection{Computational Tools}
Training of HDNNPs were performed using our own code, with the SF calculations being performed using N2P2 \cite{N2P2}.
For the CUR selection we use the code implementation from \cite{cur-implementation}.
Simulations with the trained potentials were performed in LAMMPS \cite{LAMMPS} using the ml-hdnnp plugin provided by N2P2.
As mentioned in section~\ref{sec:Datasets} we use VASP \cite{VASP} for reference \textit{ab initio} calculations.
OVITO \cite{ovito} was used for some post-processing, calculating the Radial Distribution Functions (RDFs).

\section{Results and Discussion}
\label{sec:results}

\subsection{Lennard Jones System}

\begin{figure}
  \centering
  \begin{subfigure}{\textwidth}
    \centering
    \includegraphics[width=\textwidth]{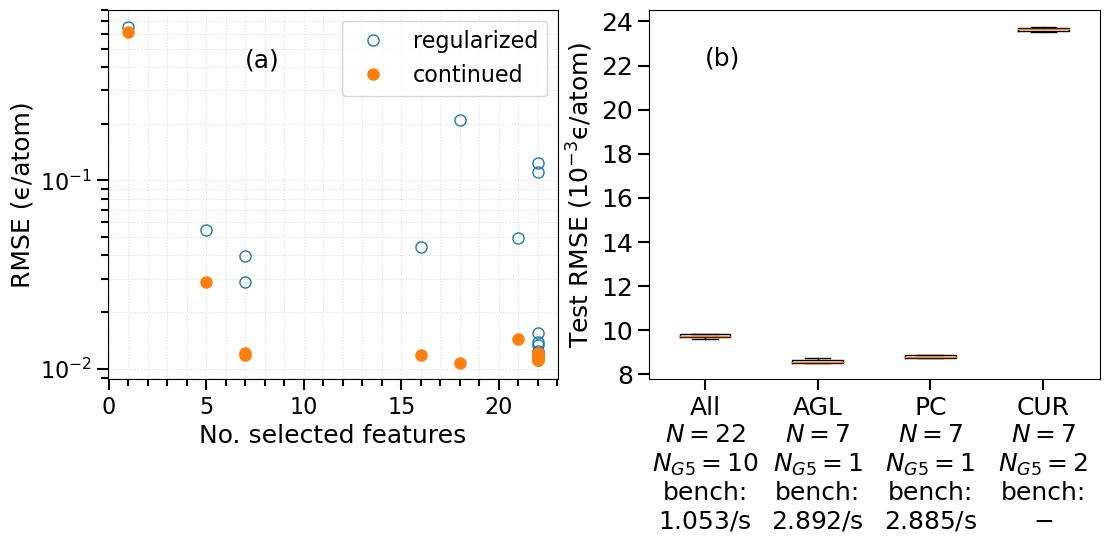}
  \end{subfigure}
  \begin{subfigure}{\textwidth}
    \centering
    \includegraphics[width=\textwidth]{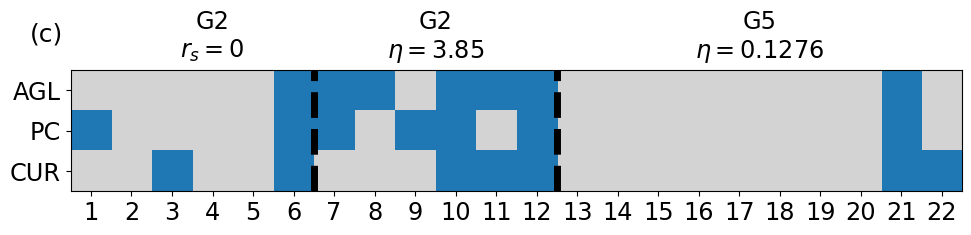}
  \end{subfigure}
  \caption{\label{fig:LJ} Selection process for LJ.
    (a) RMSE of models trained with different values of the regularization parameter $\lambda$, plotted against the number of selected features. Blue circles show the error at the end of training with the AGL penalty. Orange dots show the error after continuing training without penalty, with only selected features.
    (b) Box plot of test errors for different feature sets, with total number of features $(N)$, number of angular features $(N_{G^5})$, and timesteps per second in a benchmark simulation.
    (c) Matrix plot of selected features.
    Rows correspond to different methods, with discarded features grayed out, and selected ones in blue.
    The features are grouped into centered $G^2$, shifted $G^2$, and $G^5$.
  }
\end{figure}

As a first test of our method we apply the AGL to the LJ system, where the exact interactions are perfectly known.
In particular, they are perfectly spherically-symmetric pair interactions,
so that one might expect a feature-selection method to successfully discard
features pertaining to angular directionality.
The initial feature set contains $12$ radial SFs, $6$ of which are centered on $r_{ij}=0$ with varying widths $\eta$, with the remaining $6$ being centered on regularly spaced $r_s$ having constant width.
In addition to the radial SFs, $10$ angular ones are included, using the same wide centered radial component, with varying angular width $\zeta$ in pairs of $+1$ and $-1$ for the $\Lambda$ parameter.
All the SFs use the same cutoff radius, set to the cutoff used in the reference LJ potential, $r_c = 2.8$.
The NNP consists of two hidden layers with $10$ neurons each.

For the feature selection, we apply the AGL method described in section~\ref{sec:FS} by defining a sequence of regularization parameters $\lambda$, training an initial model with the non-adaptive GL~(\ref{eq:GL}).
This is then used to construct and retrain the model using the adaptive penalty given by (\ref{eq:AGL}).
Each of these models has its weights randomly chosen at the beginning of the training, { referred to as cold initialization}, and is trained using the ADAMW optimizer \cite{ADAMW} with learning rate set using a learning rate finder \cite{range-test}{, and a small weight decay parameter $\gamma=10^{-6}$ applied only to the internal weights so as to not interfere with the feature selection.
The batch size was fixed at $256$ configurations, and standard input normalization was used, shifting and scaling each feature to have mean $0$ and standard deviation $1$ over the training dataset.}
We let aside $10\%$ of the training data as a hold-out validation set to monitor the model performance during training for early stopping.
Crucially, for the sake of early stopping we do not monitor just the loss function, but the relevant objective function given by (\ref{eq:GL}) or (\ref{eq:AGL}){, ending training if it has not improved for $10$ epochs by more than $10^{-7}$.
  In the absence of early stopping, the training was capped at $1000$ epochs for the non-adaptive part, and $10000$ during the adaptive part.}

During training with the adaptive penalty, the weights corresponding to some of the inputs will vanish.
Following the training for each $\lambda$ we identify these weights and freeze them before continuing training without the penalty.
This is to avoid the bias that is otherwise known to occur for L1 regularized models \cite{LASSO-bias}.
Figure~\ref{fig:LJ}(a) shows the validation Root Mean Square Error (RMSE) for each model along this path, plotted against the number of selected features, both at the end of training with AGL (blue circles) and after continuing without the penalty (orange dots).
We note that the regularization introduces a noticeable overestimation of the error associated to the selected feature sets, and so continuing the training is necessary to make an informed decision on which set of selected features to choose.
In figure~\ref{fig:LJ}(a) one can observe an initial plateau in the lowest error reached during continued training when going from $22$ selected features down to $7$. We interpret this as the regime where the AGL method discards unnecessary features that lead to little decrease in performance.
Going below $7$ features, the model suffers a large increase in error, as the result of having to discard more and more important features.

Based on figure~\ref{fig:LJ}(a), we select the model with $7$ features, of which $1$ is of the angular type given by (\ref{eq:G5}).
The selected feature set is tested by training over four different random initializations, with the same training dataset, to ensure the features are not suited for just one part of the weight space.
{Unlike the models on the regularization path, in order to speed up convergence, these models were trained using the \textit{cosine annealing with warm restarts} learning rate schedule \cite{lr-schedule}.
  With this schedule the learning rate is annealed with a cosine from a large initial value to a small value ($10^{-8}$) over a number of weight updates, before resetting the learning rate to its initial value and repeating the process.
  Here the initial period of the scheduler is set to coincide with one epoch, and to double after each reset, ending training after a total of $12$ resets ($8190$ epochs).}
We likewise test the starting feature set, as well as $7$ features selected with the PC and CUR methods of \cite{imbalzano}.
The resulting test errors, evaluated on a held out test set, are presented in figure~\ref{fig:LJ}(b), together with the total number of features $N$ and the number of angular features $N_{G^5}$.
Additionally, we perform a benchmark simulation with each potential, consisting of $2048$ atoms simulated in an NVT ensemble for $10000$ timesteps.
These simulations ran on a single $2.5$ GHz Intel Cascade Lake 6248 cpu core, and the average number of simulated timesteps per second of wall time is recorded and shown in figure~\ref{fig:LJ}(b).
We note that the models trained on the features selected with CUR did not allow for a successful benchmark simulation on account of their large error, which will be discussed in more detail below.

It can be seen that there is a strong preference for radial SFs, as one would expect considering the lack of angular dependence in the reference LJ potential. Despite this, a single angular feature was selected by both the AGL and the PC filter. This is not unreasonable, since we train the LJ system with high-density configurations as reference data, where steric repulsion leads to the emergence of certain short-ranged angular order.
The features selected with CUR greatly underperform those selected with the other methods, but we note that CUR performs much better for a larger number of features \cite{imbalzano}.
CUR selected two angular features, which could allow for a better reconstruction of the atomic environment overall by taking better into account the angles, but at the cost of a reduced radial resolution.
As the CUR approach acts on the descriptors alone, it is largely incapable of knowing the lack of angular dependence of the energy in the ground truth.
It should however be mentioned that this information could still be, to some extent, indirectly available through what configurations appear in the sampled MD trajectory used to construct the dataset.

\begin{figure}
  \centering
  \includegraphics[width=0.48\textwidth]{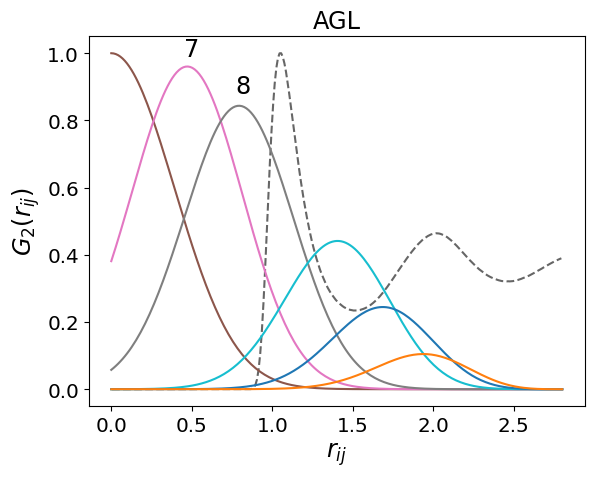}
  \includegraphics[width=0.48\textwidth]{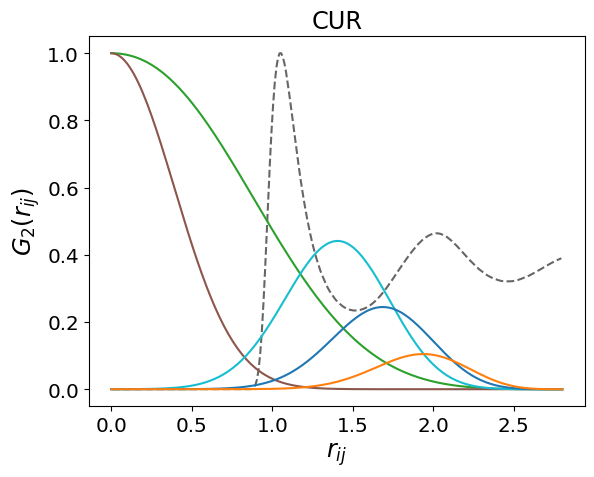}
  \caption{\label{fig:LJ-G2}Radial symmetry functions ($G_2$) for LJ, visualized for a single atom pair, selected by AGL (left), and CUR (right). RDF of reference system included for comparison (dashed line). Features number $7$ and $8$ from figure~\ref{fig:LJ}(c) are marked.}
\end{figure}

{
To better illustrate the differences between the feature selection methods, we show in figure~\ref{fig:LJ}(c) a matrix representing the features selected by each method.
The $G^2$ SFs selected by AGL and CUR are also plotted in figure~\ref{fig:LJ-G2}, along with the Radial Distribution Function (RDF) extracted from one of the reference simulations.
Of note is that CUR discarded three consecutive shifted radial SFs in a regime where the other methods kept at least one.
This raises the question of whether adding one of these SFs to the CUR features would recover a good performance.
In order to test this, we create two new sets by adding to the CUR features one of the shifted radial SFs selected by AGL but discarded by CUR, marked $7$ and $8$ in figure~\ref{fig:LJ-G2}.
Adding feature number $8$ reduced the test RMSE to $18.4\times10^{-3}\epsilon$/atom, which is a modest improvement, but still nowhere near the performance of the other sets.
Instead, adding feature number $7$ lowers the test RMSE to $9.40\times10^{-3}\epsilon$/atom, a clear indication that this is indeed a vitally important feature for this system that the CUR method failed to detect.}
{With this feature added, the resulting model also allowed for stable simulations to be performed.}

\subsection{Aluminium}

\begin{figure}
  \centering
  \begin{subfigure}{\textwidth}
    \includegraphics[width=\textwidth]{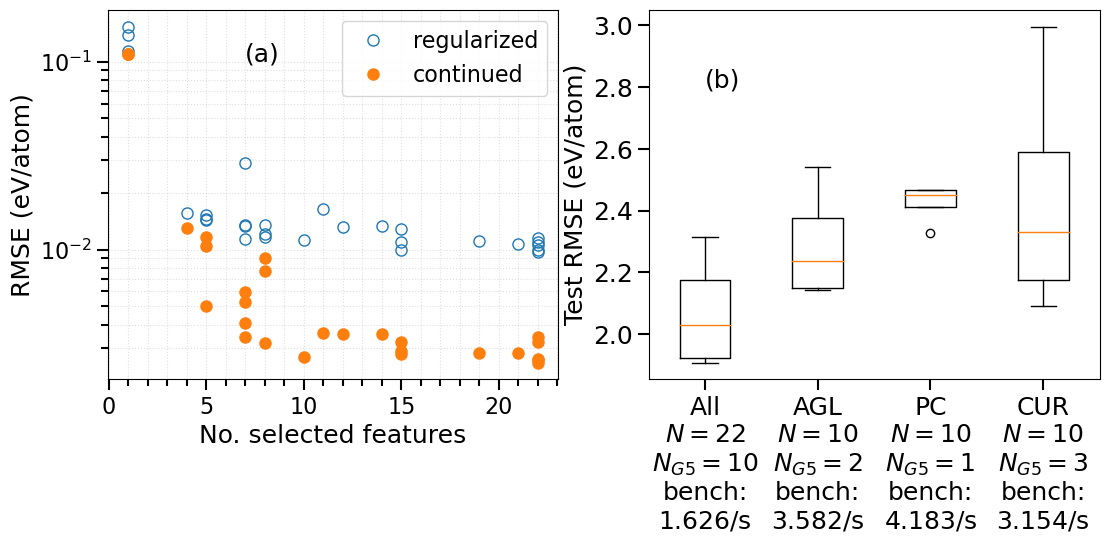}
  \end{subfigure}
  \begin{subfigure}{\textwidth}
    \centering
    \includegraphics[width=\textwidth]{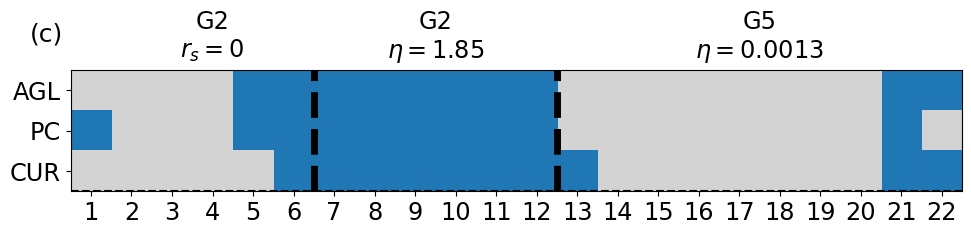}
  \end{subfigure}
  \caption{\label{fig:Al} Selection process for Al.
    (a) RMSE of models trained with different values of the regularization parameter $\lambda$, plotted against the number of selected features. Blue circles show the error at the end of training with the AGL penalty. Orange dots show the error after continuing training without penalty, with only selected features.
    (b) Box plot of test errors for different feature sets, with total number of features $(N)$, number of angular features $(N_{G^5})$, and timesteps per second in a benchmark simulation.
    (c) Matrix plot of selected features.
      Rows correspond to different methods, with discarded features grayed out, and selected ones in blue.
  The features are grouped into centered $G^2$, shifted $G^2$, and $G^5$.}
\end{figure}

\begin{figure}
  \centering
  \begin{subfigure}{0.5\textwidth}
    \includegraphics[width=\textwidth]{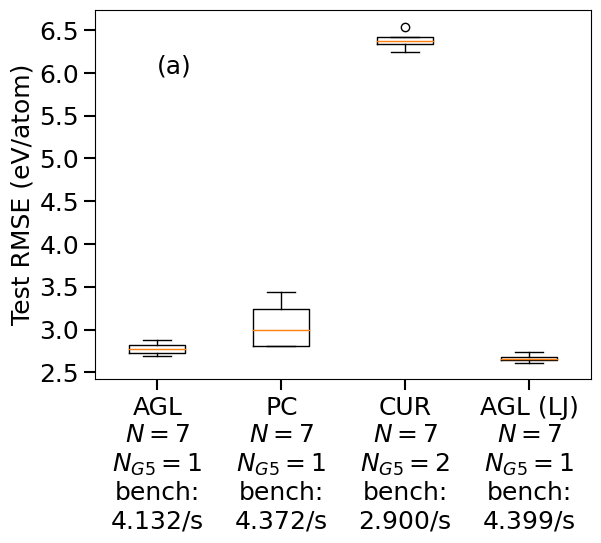}
  \end{subfigure}
  \begin{subfigure}{\textwidth}
    \centering
    \includegraphics[width=\textwidth]{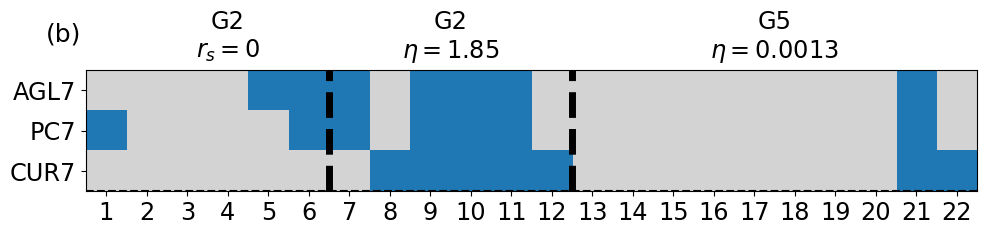}
  \end{subfigure}
  \caption{\label{fig:Al-7}
    (a) Box plot of test errors for different feature sets, with total number of features $(N)$, number of angular features $(N_{G^5})$, and timesteps per second in a benchmark simulation.
    (b) Matrix plot of features selected for Al, with 7 features selected.
      Rows correspond to different methods, with discarded features grayed out, and selected ones in blue.
  The features are grouped into centered $G^2$, shifted $G^2$, and $G^5$.}
\end{figure}

To test the method in a more practical setting, we turn to the case of Al.
The SF parameters and network architecture is chosen as in \cite{Jakse-2023-Al}.
We proceed as for LJ, training a sequence of models on increasing values of $\lambda$, using cold initialization, continuing the training after selecting the features.
The resulting validation errors are plotted against the number of selected features in figure~\ref{fig:Al}(a).
We find 10 features to be a good compromise between few features and low error.
The set is again evaluated by training a set of four models on the selected features, with different initialization, likewise for the staring features and features selected with CUR and PC.
The test errors are shown in figure~\ref{fig:Al}(b), along with the number of angular features selected, and number of timesteps per second in a benchmark simulation identical to the one for LJ.
We see a significant increase in computational speed for the feature-selected potential, at a relatively small increase in error.
For this system, CUR and PC seem to perform equivalently.
In particular the CUR features perform much better than in the LJ case, presumably because it is asked to select more features and so the method is not forced to compromise on the radial resolution.
The features selected with AGL, on average, outperform those chosen by the filters, although there is not a large difference in this case, especially considering the deviations.

{The feature sets are visualized in figure~\ref{fig:Al}(c).
  We observe, somewhat different from the LJ case, a great overlap between the methods, and presumably the one or two features that differ between each set are not enough to cause a significant difference in the test error.
  In particular we notice that each model selected each shifted radial SF.
  Feature number 6 in figure~\ref{fig:Al}(c), being also selected by each model, is identical to the shifted ones, but centered on $r_{ij}=0$.
  Taken together these features can be argued to cover the entire range of interatomic distances up to the cutoff radius, allowing for a rough representation of the RDF.
  This preference for shifted radial SFs has also been indicated elsewhere in the literature \cite{wACSF}.}

Like in the case of LJ, there is here a preference towards radial features, with only two angular ones being chosen.
We suggest a physical explanation for this {preference for radial features}, noting the tendency of Al to adopt a close-packed short range order and to maximize the number of nearest neighbors, due to the weakly directional sp bonding type electronic structure.

{
  While the 10 features selected are a sensible choice, based on the training errors reported in figure~\ref{fig:Al}(a), the threshold is not rigorous. From the RMSE values obtained, a selection of 8 or even only 7 features could also be argued for.
Hence we also show in figure~\ref{fig:Al-7}(a) errors of models trained with the best performing set of 7 features, as well as corresponding sets selected with PC and CUR.
  A noticeable increase in the test error is observed, with only a modest improvement in benchmark performance primarily due to the additional discarded angular SF.
  We note that the CUR features show a significant reduction in performance, reminiscent of what was observed for LJ. In the present case, this is presumably due to the deselection of both features number $6$ and $7$ by CUR, seen if figure~\ref{fig:Al-7}(b).
  As this is the same number of features selected for LJ, one can also compare the two sets of features.
  We note that the selected angular feature is the same in both cases.
  As a test, we train a model for Al with the set of features selected by AGL for LJ, denoted AGL (LJ) in figure~\ref{fig:Al-7}(a).
  Interestingly the LJ set seems to slightly outperform the other $7$ features selected by AGL.
}

\subsection{Boron}

\begin{figure}
  \centering
  \begin{subfigure}{\textwidth}
    \includegraphics[width=\textwidth]{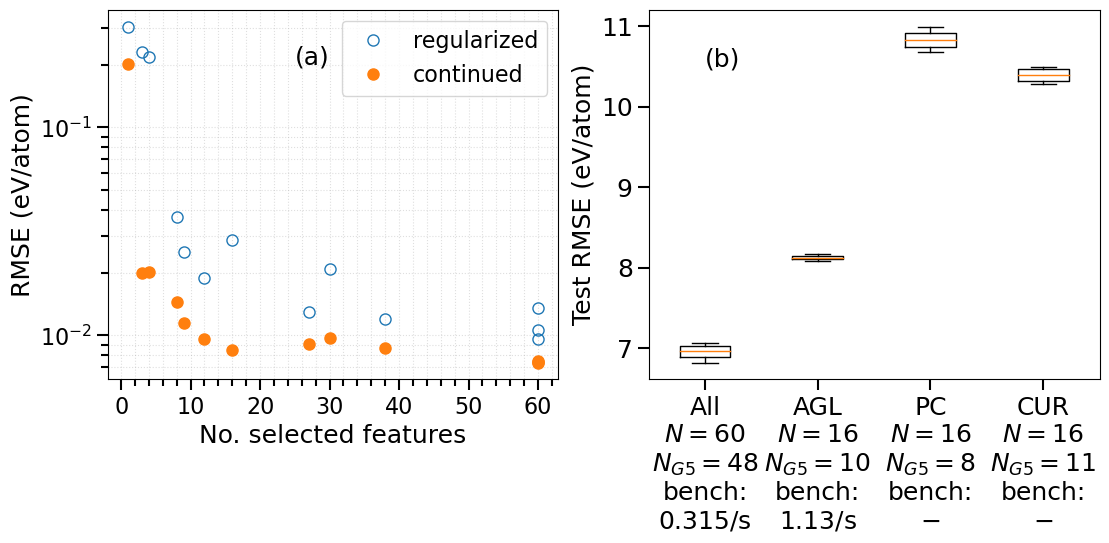}
  \end{subfigure}
  \begin{subfigure}{\textwidth}
    \includegraphics[width=\textwidth]{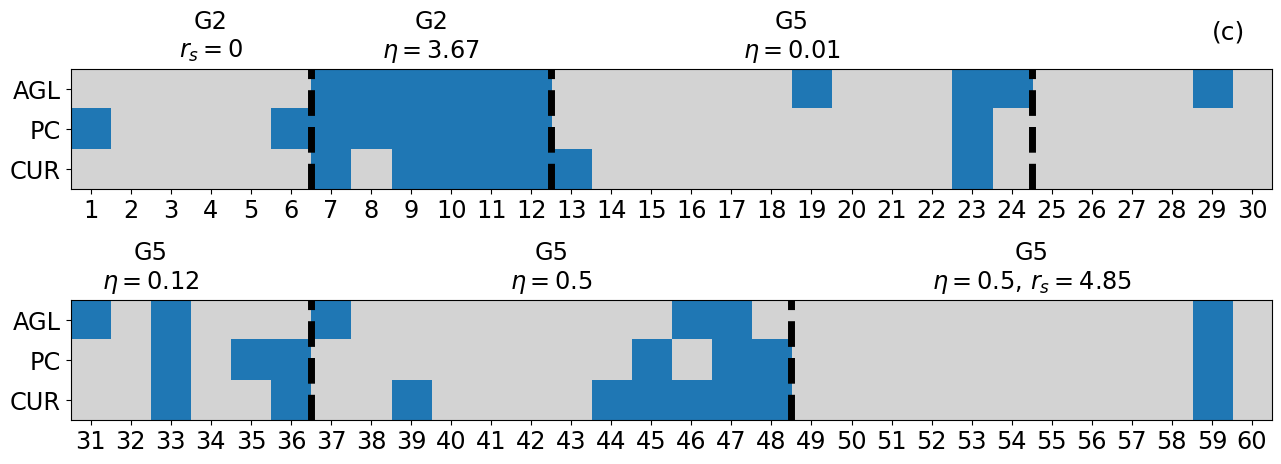}
  \end{subfigure}
  \caption{\label{fig:B} Selection process for B.
    (a) RMSE of models trained with different values of the regularization parameter $\lambda$, plotted against the number of selected features. Blue circles show the error at the end of training with the AGL penalty. Orange dots show the error after continuing training without penalty, with only selected features.
    (b) Box plot of test errors for different feature sets, with total number of features $(N)$, number of angular features $(N_{G^5})$, and timesteps per second in a benchmark simulation.
    (c) Matrix plot of selected features. Rows correspond to different methods, with discarded features grayed out, and selected ones in blue.
  }
\end{figure}

We turn to boron as a stringent test system.
Due to the complicated structure of boron, induced by strong covalent directional bonding \cite{beta-B-review, B-review-2009, Jakse-2014-B}, we expect this to be a significantly more difficult task, and to require a more complex set of features compared to Al and LJ.
For our initial set of descriptors we use a set of 12 radial SFs, and 48 angular SFs, with a cutoff of $5.3\;\AA$ corresponding roughly to the outer edge of the third neighbor shell.
This relatively wide cutoff was chosen in order to hopefully be able to more adequately take into account the medium-range structure known to appear in boron, primarily the open icosahedra and the bonds between them \cite{Jakse-2014-B}.
{Furthermore, to allow for a potentially more complex mapping we use a larger network than for LJ and Al, with two layers of $25$ hidden nodes each, providing a slight improvement in error compared to smaller network sizes.}

As for the previous systems, figure~\ref{fig:B}(a) shows the validation RMSE as a function of the selected features.
In this case the best-performing model, apart from the one with the full set of features, is for $16$ features.
We select these 16 features, and again train a set of four models to test, with the results shown in figure~\ref{fig:B}(b).
In this case we not only selected a larger number of features, but the majority of features selected were of the angular type.
Unlike in the previous cases, we also observe an inability of the filter methods to adequately select features for this system, with a significant increase in error for the set selected with PC and CUR.
In fact, we were unable to perform even a benchmark simulation using the models trained on the PC and CUR sets, with the simulations becoming unstable.
For the AGL set there is a noticeable increase in the error compared to the full set of features, but this comes with a significant improvement in the computational performance of the potential.

{
  The selected features for each set is shown in figure~\ref{fig:B}(c).
  We see, as for Al, that the shifted radial SFs are seemingly the most important radial ones.
  For the angular features the picture is, however, not very clear.
In particular it is not \textit{a priori} evident why the features selected by CUR and PC
lead to worse performance than the ones selected by AGL.
  Like for the LJ case we tried adding individual features from the set selected by AGL to the set selected by CUR, to see if this improves the performance.
  Adding features number $29$ and $31$ to the CUR set changes the error to $8.95$ meV/atom, and $9.68$ meV/atom respectively, neither of which allowed for stable simulation.
  A model with both of these added did not reduce the error any further, and was likewise unstable.
  None of the other features we tried adding managed to reduce the error below $10$ meV/atom.
}

\subsection{Validation of the MLIP models}\label{sec:validation}

\begin{table}
  \centering
  \caption{\label{tab:diffusion}
  Comparison of diffusion constants $D$ predicted with different sets of $N$ features to \textit{ab initio}}
  \begin{tabular}{c|cc|cc|cc}\hline
    & \multicolumn{2}{c|}{LJ} & \multicolumn{2}{c|}{Al} & \multicolumn{2}{c}{B}\\
    method & \phantom{m}$N$\phantom{m} & $D$ ($\sigma^2/\tau$) & \phantom{m}$N$\phantom{m} & $D$ (\AA$^2$/s) & \phantom{m}$N$\phantom{m} & $D$ (\AA$^2$/s)\\\hline
    AIMD & - & $0.0625\pm0.0054$ & - & $1.6080$ & - & $1.8176$\\
    - & $22$ & $0.0603\pm0.0040$ & $22$ & $1.5483 \pm 0.0845$ & $60$ & $1.8265 \pm 0.0880$\\
    AGL & $7$ & $0.0631\pm0.0029$ & $10$ & $1.6431 \pm 0.0620$ & $16$ & $1.8634 \pm 0.0875$\\
    PC & $7$ & $0.0633\pm0.0039$ & $10$ & $1.5843 \pm 0.0607$ & $16$ & -\\
    CUR & $7$ & - & $10$ & $1.5813 \pm 0.0920$ & $16$ & -\\\hline
  \end{tabular}
\end{table}

\begin{figure}
  \centering
  \includegraphics[width=0.32\textwidth]{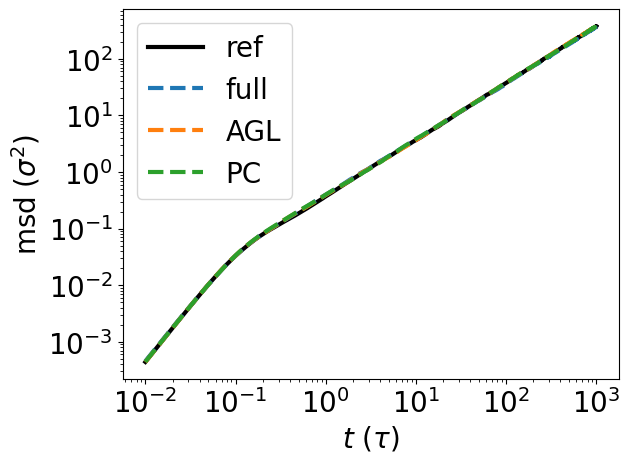}
  \includegraphics[width=0.32\textwidth]{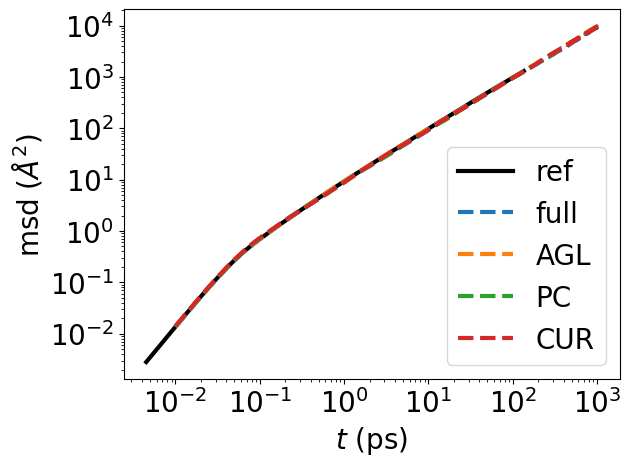}
  \includegraphics[width=0.32\textwidth]{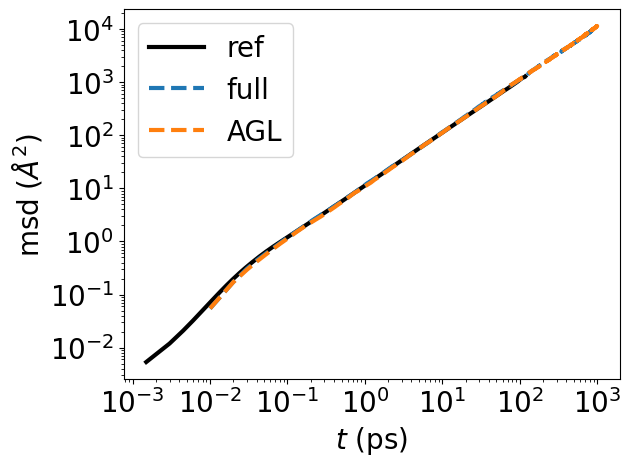}
  \caption{\label{fig:msd}MSDs for different feature sets for LJ (left), Al (center), and B (right). For each feature set the best test model is used.}
\end{figure}

While looking at the RMSE of the models on a held-out set of test configuration is useful, the true test of the quality of a MLIP is in simulations and the accurate prediction of physical quantities.
For each set of features we pick out the model with the best test error and perform an NVT simulation, aiming to obtain the diffusion constant for comparison to \textit{ab initio}, and the reference potential in the case of LJ.
Each simulation uses a box of 256 atoms, in order to match the finite size effect in the reference systems.
The temperature is $1500\; K$ for Al, $2600\; K$ for B, and $1.5\; k_{B}/\epsilon$ for LJ, and in each case the simulation box is chosen such that the density is the same as in the respective reference system.
For Al and LJ the system is initialized in an fcc crystal configuration and evolved until it melts, while for B we initialize with a liquid configuration taken from the AIMD dataset.
Each simulation consists of 10 measurements of 1M timesteps, each preceded by a $10^5$ timestep equilibration after randomly reassigning the atomic velocities.
{Over these trajectories we calculate the Mean Square Displacement (MSD), shown in figure~\ref{fig:msd} for each system.
  In none of the cases do we observe any clear difference induced by the featureset.
  To give a clearer picture, the diffusion constants are extracted from the MSD using the Einstein relations.}
The predicted diffusion constants for each feature set, and reference values, are shown in table~\ref{tab:diffusion}.
In all cases the diffusion constants are within acceptable bounds of the reference values, with no discernible effect resulting from the difference in number of variables.
We note that, as with the benchmark, the models trained for B with the filter methods did not allow for successful simulations, but were too unstable. The same held true for the features selected for LJ using the CUR method.

\begin{figure}
  \centering
  \includegraphics[width=0.32\textwidth]{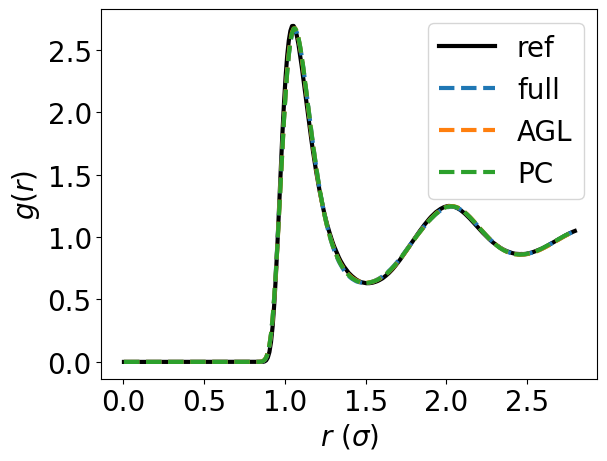}
  \includegraphics[width=0.32\textwidth]{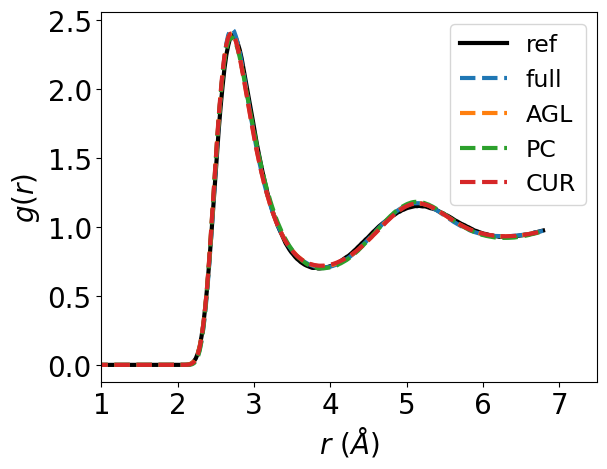}
  \includegraphics[width=0.32\textwidth]{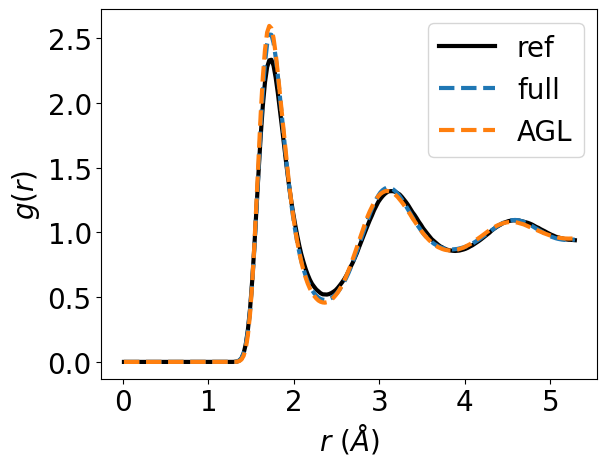}
  \caption{\label{fig:rdf}RDFs for different feature sets for LJ (left), Al (center), and B (right). For each feature set the best test model is used.}
\end{figure}

{
From these simulations we also extract the RDF, shown for each system in figure~\ref{fig:rdf}.
One point that should be stressed here is that our aim is to evaluate the feature selection, rather than how well any of the models reproduce the AIMD reference system results.
For both the Al and LJ cases we observe very little difference between the different NNP models, as both the initial large feature set and also the reduced sets following feature selection reproduce the AIMD results fairly well.
In the case of B, already the initial large feature set turns out to be not powerful enough to reproduce the boron RDF faithfully. But the feature selection by AGL does not deteriorate the agreement further, indicating that no significant performance is lost -- the feature selection can be only as good as the initial starting point. The failure to reproduce the AIMD RDF emphasizes that boron is a challenging system for the training based on Behler-Parrinello SFs and potential energies as targets. Irrespective of this, the agreement with the AIMD MSD is very good also for the reduced feature set. We rationalize this as a result of the dynamics in boron being not predominantly determined by the radial structure encoded in the angle-averaged RDF.}
{
This additionally points to the possibility of the standard BP SFs being not well suited for this system, as previously suggested in the literature \cite{B-atomic-structure}.}

\subsection{Confounding Features}
\begin{figure}
  \centering
  \includegraphics[width=\textwidth]{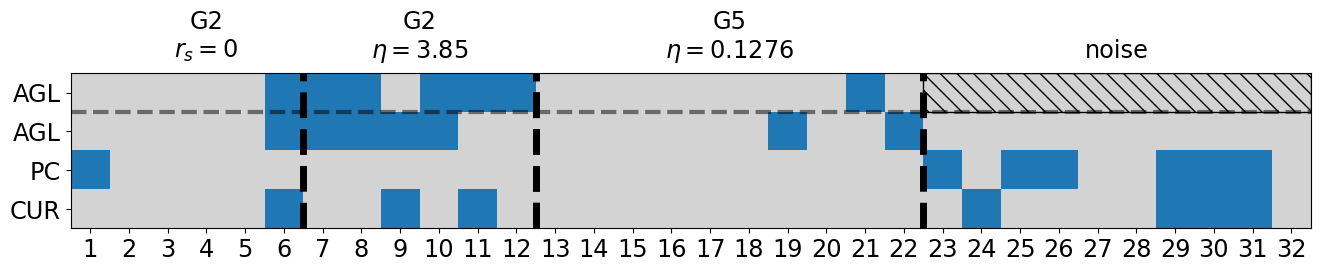}
  \caption{\label{fig:noise-set}
    Matrix plot of selected features for LJ, with random noise features.
    Rows correspond to different methods, with discarded features grayed out, and selected ones in blue.
    Top row, above the dashed line, is from figure~\ref{fig:LJ}~(c), included for reference.}
\end{figure}

Filter methods such as PC and CUR aim to reduce the number of features by looking for subsets that minimize the overlap between those features that are kept. However, this makes them potentially vulnerable to confounding features that are uncorrelated to the relevant input, but by themselves irrelevant. This requires the initial selection of features one starts with to be carefully chosen, in order to minimize irrelevant input. However, in a system with complex structure this might not be obvious to achieve. We demonstrate in the following, that AGL performs much better in the presence of irrelevant input.

For this purpose we return to the LJ system, modifying the starting featureset by adding $10$ new features consisting of random noise drawn independently from a set of Gaussian distributions, with means and variances chosen to mimic those of the real features.
We note that these fake features were sampled once for each atom and configuration, and as such the values do not vary between epochs.
To ensure these fake features are nonnegative, like the real ones, we only work with their absolute values.
While the situation considered here is a rather implausible one to occur in a practical setting, where features are unlikely to be truly uncorrelated to the potential energy, it could potentially have implications in situations where there is noise in the training dataset.

Having nothing to do with the real data generating process, these features are truly independent from the other features as well as the target energy.
Ideally these features should be discarded, but as they are independent from the real features as well as each other, we expect that neither the PC nor CUR should be able to correctly discard them.
This is indeed the case, as illustrated in figure~\ref{fig:noise-set}, showing the features selected by AGL, PC, and CUR, as well as for comparison, the set selected by AGL in the absence of fakes.
The PC method clearly did not succeed, as beyond the manually selected feature it only picked out fake features.
With CUR we selected some real features, indicating that the method might be more robust compared to the PC in this regard, but still it selected more fakes than real features.
In contrast to the filters, the AGL managed to discard the fakes, and select a set of features.
And interesting observation is that the set selected by the AGL is slightly different to that selected in the absence of fakes.
In fact, the error obtained on this set was $6.97\times10^{-3}$, below that of the set selected in absence of fakes.
This is reminiscent of machine learning methods where the deliberate addition of noise helps increasing the performance in training.

\section{Conclusion and Outlook}
\label{sec:conclusion}

{
We have applied the AGL as an embedded feature selection method for choosing atomic features in HDNNPs.
This allows for selecting features as part of the training process, taking into account the action of the features in the resulting potential during the selection.
In order to evaluate the method we have compared it to previously used unsupervised filter methods that take only into account the features themselves, aiming to minimize redundancy in the description of the local atomic environment.
We find that for three test systems, ranging from a simple LJ system, to the highly complicated and directional boron system, that the AGL manages to perform as good as, or better than the other methods.
This we consider the main outcome of this work.
By utilizing a method that takes into account the NNP predictions, we can reduce the number of atomic features further than methods taking only into account the features themselves.

While we have applied our method to training on only energies, the next step would be to apply the method to the more common setting of fitting also forces during training.
A natural question in this case is whether the inclusion of forces changes the features that are selected.
It would also be a natural direction to use the method for different types of descriptors.
Although the BP SFs are largely in use, and have seen plenty of success, since their introduction many other alternative descriptors have been developed.
This is especially relevant considering the difficulty of even our full set of features to reproduce the RDF of boron, which could be an indication that the SFs are not ideally suited for this system.
One can further consider applying the method to multicomponent systems, where a traditional SF approach sees a combinatorial increase in the number of features, which could potentially be counteracted by feature selection.
In view of recent concerns regarding the stability of MLIPs \cite{stability-2023}, it would also be interesting to study the extent to which input dimensionality affects the stability of models, and whether this can be alleviated by careful feature selection, or indeed regularization in general.

}

\section*{Code and Data Availability}
The authors will make the data of this study available upon reasonable request.
The code used in this article is available at \url{https://github.com/JohannesSandberg/HDNNP-AGL}

\section*{Acknowledgments}

We acknowledge the CINES and IDRIS under Project No. INP2227/72914, as well as CIMENT/GRICAD for computational resources. This work was performed within the framework of the Centre of Excellence of Multifunctional Architectured Materials “CEMAM” ANR-10-LABX-44-01 funded by the “Investments for the Future” Program. This work has been partially supported by MIAI@Grenoble Alpes (ANR-19-P3IA-0003).
JS acknowledges funding from the German Academic Exchange Service through DLR-DAAD fellowship grant number 509.
We thank Gerhard Jung for suggesting tests with random features.

\section*{References}
\bibliographystyle{unsrt}
\bibliography{references.bib}

\begin{thebibliography}{10}

\bibitem{behler-2016-perspective}
J{\"o}rg Behler.
\newblock Perspective: Machine learning potentials for atomistic simulations.
\newblock {\em The Journal of chemical physics}, 145(17), 2016.

\bibitem{behler-2022-review}
Emir Kocer, Tsz~Wai Ko, and J{\"o}rg Behler.
\newblock Neural network potentials: A concise overview of methods.
\newblock {\em Annual review of physical chemistry}, 73:163--186, 2022.

\bibitem{Schmidt-2019-review}
Jonathan Schmidt, M{\'a}rio~RG Marques, Silvana Botti, and Miguel~AL Marques.
\newblock Recent advances and applications of machine learning in solid-state
  materials science.
\newblock {\em npj Computational Materials}, 5(1):83, 2019.

\bibitem{Choudhary-2022-review}
Kamal Choudhary, Brian DeCost, Chi Chen, Anubhav Jain, Francesca Tavazza, Ryan
  Cohn, Cheol~Woo Park, Alok Choudhary, Ankit Agrawal, Simon~JL Billinge,
  et~al.
\newblock Recent advances and applications of deep learning methods in
  materials science.
\newblock {\em npj Computational Materials}, 8(1):59, 2022.

\bibitem{Hafner-2008}
J{\"u}rgen Hafner.
\newblock Ab-initio simulations of materials using vasp: Density-functional
  theory and beyond.
\newblock {\em Journal of computational chemistry}, 29(13):2044--2078, 2008.

\bibitem{EAM}
Murray~S Daw and Michael~I Baskes.
\newblock Embedded-atom method: Derivation and application to impurities,
  surfaces, and other defects in metals.
\newblock {\em Physical Review B}, 29(12):6443, 1984.

\bibitem{MEAM}
Michael~I Baskes.
\newblock Modified embedded-atom potentials for cubic materials and impurities.
\newblock {\em Physical review B}, 46(5):2727, 1992.

\bibitem{ReaxFF}
Adri~CT Van~Duin, Siddharth Dasgupta, Francois Lorant, and William~A Goddard.
\newblock Reaxff: a reactive force field for hydrocarbons.
\newblock {\em The Journal of Physical Chemistry A}, 105(41):9396--9409, 2001.

\bibitem{Jakse-2023-Al}
Noel Jakse, Johannes Sandberg, Leon~F Granz, Anthony Saliou, Philippe Jarry,
  Emilie Devijver, Thomas Voigtmann, J{\"u}rgen Horbach, and Andreas Meyer.
\newblock Machine learning interatomic potentials for aluminium: application to
  solidification phenomena.
\newblock {\em Journal of Physics: Condensed Matter}, 51(3):035402, 2022.

\bibitem{nucleation-H2O}
Pablo~M Piaggi, Jack Weis, Athanassios~Z Panagiotopoulos, Pablo~G Debenedetti,
  and Roberto Car.
\newblock Homogeneous ice nucleation in an ab initio machine-learning model of
  water.
\newblock {\em Proceedings of the National Academy of Sciences},
  119(33):e2207294119, 2022.

\bibitem{Al-Cu-NNP-2020}
Daniel Marchand, Abhinav Jain, Albert Glensk, and WA~Curtin.
\newblock Machine learning for metallurgy i. a neural-network potential for
  al-cu.
\newblock {\em Physical review materials}, 4(10):103601, 2020.

\bibitem{Al-Mg-Si-NNP-2021}
Abhinav~CP Jain, Daniel Marchand, Albert Glensk, M~Ceriotti, and WA~Curtin.
\newblock Machine learning for metallurgy iii: A neural network potential for
  al-mg-si.
\newblock {\em Physical Review Materials}, 5(5):053805, 2021.

\bibitem{Al-Cu-Mg-Zn-NNP-2022}
Daniel Marchand and WA~Curtin.
\newblock Machine learning for metallurgy iv: A neural network potential for
  al-cu-mg and al-cu-mg-zn.
\newblock {\em Physical Review Materials}, 6(5):053803, 2022.

\bibitem{Li-Si-Amorphous-NNP}
Nongnuch Artrith, Alexander Urban, and Gerbrand Ceder.
\newblock Constructing first-principles phase diagrams of amorphous lixsi using
  machine-learning-assisted sampling with an evolutionary algorithm.
\newblock {\em The Journal of chemical physics}, 148(24), 2018.

\bibitem{Li3PO4-Amorphous-NNP}
Wenwen Li, Yasunobu Ando, Emi Minamitani, and Satoshi Watanabe.
\newblock Study of li atom diffusion in amorphous li3po4 with neural network
  potential.
\newblock {\em The Journal of chemical physics}, 147(21), 2017.

\bibitem{SNAP}
Aidan~P Thompson, Laura~P Swiler, Christian~R Trott, Stephen~M Foiles, and
  Garritt~J Tucker.
\newblock Spectral neighbor analysis method for automated generation of
  quantum-accurate interatomic potentials.
\newblock {\em Journal of Computational Physics}, 285:316--330, 2015.

\bibitem{GAP}
Albert~P Bart{\'o}k, Mike~C Payne, Risi Kondor, and G{\'a}bor Cs{\'a}nyi.
\newblock Gaussian approximation potentials: The accuracy of quantum mechanics,
  without the electrons.
\newblock {\em Physical review letters}, 104(13):136403, 2010.

\bibitem{Gaussian-Process-Regression}
Volker~L Deringer, Albert~P Bart{\'o}k, Noam Bernstein, David~M Wilkins,
  Michele Ceriotti, and G{\'a}bor Cs{\'a}nyi.
\newblock Gaussian process regression for materials and molecules.
\newblock {\em Chemical Reviews}, 121(16):10073--10141, 2021.

\bibitem{DeepMD}
Linfeng Zhang, Jiequn Han, Han Wang, Roberto Car, and EJPRL Weinan.
\newblock Deep potential molecular dynamics: a scalable model with the accuracy
  of quantum mechanics.
\newblock {\em Physical review letters}, 120(14):143001, 2018.

\bibitem{behler-parrinello-2007}
J{\"o}rg Behler and Michele Parrinello.
\newblock Generalized neural-network representation of high-dimensional
  potential-energy surfaces.
\newblock {\em Physical review letters}, 98(14):146401, 2007.

\bibitem{SchNET}
Kristof~T Sch{\"u}tt, Huziel~E Sauceda, P-J Kindermans, Alexandre Tkatchenko,
  and K-R M{\"u}ller.
\newblock Schnet--a deep learning architecture for molecules and materials.
\newblock {\em The Journal of Chemical Physics}, 148(24), 2018.

\bibitem{NequIP}
Simon Batzner, Albert Musaelian, Lixin Sun, Mario Geiger, Jonathan~P Mailoa,
  Mordechai Kornbluth, Nicola Molinari, Tess~E Smidt, and Boris Kozinsky.
\newblock E (3)-equivariant graph neural networks for data-efficient and
  accurate interatomic potentials.
\newblock {\em Nature communications}, 13(1):2453, 2022.

\bibitem{4-gens}
Jorg Behler.
\newblock Four generations of high-dimensional neural network potentials.
\newblock {\em Chemical Reviews}, 121(16):10037--10072, 2021.

\bibitem{behler-perspective-2016}
J{\"o}rg Behler.
\newblock Perspective: Machine learning potentials for atomistic simulations.
\newblock {\em The Journal of chemical physics}, 145(17), 2016.

\bibitem{wACSF}
Michael Gastegger, Ludwig Schwiedrzik, Marius Bittermann, Florian Berzsenyi,
  and Philipp Marquetand.
\newblock wacsf—weighted atom-centered symmetry functions as descriptors in
  machine learning potentials.
\newblock {\em The Journal of chemical physics}, 148(24), 2018.

\bibitem{FS-review-2014}
Girish Chandrashekar and Ferat Sahin.
\newblock A survey on feature selection methods.
\newblock {\em Computers \& Electrical Engineering}, 40(1):16--28, 2014.

\bibitem{imbalzano}
Giulio Imbalzano, Andrea Anelli, Daniele Giofr{\'e}, Sinja Klees, J{\"o}rg
  Behler, and Michele Ceriotti.
\newblock Automatic selection of atomic fingerprints and reference
  configurations for machine-learning potentials.
\newblock {\em The Journal of chemical physics}, 148(24), 2018.

\bibitem{CUR}
Michael~W Mahoney and Petros Drineas.
\newblock Cur matrix decompositions for improved data analysis.
\newblock {\em Proceedings of the National Academy of Sciences},
  106(3):697--702, 2009.

\bibitem{LassoNET}
Ismael Lemhadri, Feng Ruan, and Rob Tibshirani.
\newblock Lassonet: Neural networks with feature sparsity.
\newblock In {\em International conference on artificial intelligence and
  statistics}, pages 10--18. PMLR, 2021.

\bibitem{LASSO}
Robert Tibshirani.
\newblock Regression shrinkage and selection via the lasso.
\newblock {\em Journal of the Royal Statistical Society Series B: Statistical
  Methodology}, 58(1):267--288, 1996.

\bibitem{Seko-LASSO}
Atsuto Seko, Akira Takahashi, and Isao Tanaka.
\newblock Sparse representation for a potential energy surface.
\newblock {\em Physical Review B}, 90(2):024101, 2014.

\bibitem{Seko-ElasticNet}
Atsuto Seko, Akira Takahashi, and Isao Tanaka.
\newblock First-principles interatomic potentials for ten elemental metals via
  compressed sensing.
\newblock {\em Physical Review B}, 92(5):054113, 2015.

\bibitem{Ghiringhelli-2017}
Luca~M Ghiringhelli, Jan Vybiral, Emre Ahmetcik, Runhai Ouyang, Sergey~V
  Levchenko, Claudia Draxl, and Matthias Scheffler.
\newblock Learning physical descriptors for materials science by compressed
  sensing.
\newblock {\em New Journal of Physics}, 19(2):023017, 2017.

\bibitem{SISSO}
Runhai Ouyang, Stefano Curtarolo, Emre Ahmetcik, Matthias Scheffler, and Luca~M
  Ghiringhelli.
\newblock Sisso: A compressed-sensing method for identifying the best
  low-dimensional descriptor in an immensity of offered candidates.
\newblock {\em Physical Review Materials}, 2(8):083802, 2018.

\bibitem{original-GL}
Ming Yuan and Yi~Lin.
\newblock Model selection and estimation in regression with grouped variables.
\newblock {\em Journal of the Royal Statistical Society Series B: Statistical
  Methodology}, 68(1):49--67, 2006.

\bibitem{AGL}
Vu~C Dinh and Lam~S Ho.
\newblock Consistent feature selection for analytic deep neural networks.
\newblock {\em Advances in Neural Information Processing Systems},
  33:2420--2431, 2020.

\bibitem{beta-B-review}
Tadashi Ogitsu, Eric Schwegler, and Giulia Galli.
\newblock $\beta$-rhombohedral boron: At the crossroads of the chemistry of
  boron and the physics of frustration.
\newblock {\em Chemical reviews}, 113(5):3425--3449, 2013.

\bibitem{B-review-2009}
Barbara Albert and Harald Hillebrecht.
\newblock Boron: elementary challenge for experimenters and theoreticians.
\newblock {\em Angewandte Chemie International Edition}, 48(46):8640--8668,
  2009.

\bibitem{LAMMPS}
Aidan~P Thompson, H~Metin Aktulga, Richard Berger, Dan~S Bolintineanu,
  W~Michael Brown, Paul~S Crozier, Pieter~J in't Veld, Axel Kohlmeyer, Stan~G
  Moore, Trung~Dac Nguyen, et~al.
\newblock Lammps-a flexible simulation tool for particle-based materials
  modeling at the atomic, meso, and continuum scales.
\newblock {\em Computer Physics Communications}, 271:108171, 2022.

\bibitem{LJ-phase-diagram}
Andrew~J Schultz and David~A Kofke.
\newblock Comprehensive high-precision high-accuracy equation of state and
  coexistence properties for classical lennard-jones crystals and
  low-temperature fluid phases.
\newblock {\em The Journal of Chemical Physics}, 149(20), 2018.

\bibitem{VASP}
Georg Kresse and J{\"u}rgen Furthm{\"u}ller.
\newblock Efficiency of ab-initio total energy calculations for metals and
  semiconductors using a plane-wave basis set.
\newblock {\em Computational materials science}, 6(1):15--50, 1996.

\bibitem{LDA}
John~P Perdew and Alex Zunger.
\newblock Self-interaction correction to density-functional approximations for
  many-electron systems.
\newblock {\em Physical Review B}, 23(10):5048, 1981.

\bibitem{Jakse-2014-B}
N~Jakse and A~Pasturel.
\newblock Interplay between the structure and dynamics in liquid and
  undercooled boron: An ab initio molecular dynamics simulation study.
\newblock {\em The Journal of Chemical Physics}, 141(23), 2014.

\bibitem{materials-project}
Anubhav Jain, Shyue~Ping Ong, Geoffroy Hautier, Wei Chen, William~Davidson
  Richards, Stephen Dacek, Shreyas Cholia, Dan Gunter, David Skinner, Gerbrand
  Ceder, et~al.
\newblock Commentary: The materials project: A materials genome approach to
  accelerating materials innovation.
\newblock {\em APL materials}, 1(1), 2013.

\bibitem{Perdew-Wang-GGA}
Yue Wang and John~P Perdew.
\newblock Correlation hole of the spin-polarized electron gas, with exact
  small-wave-vector and high-density scaling.
\newblock {\em Physical Review B}, 44(24):13298, 1991.

\bibitem{behler-tutorial-2015}
J{\"o}rg Behler.
\newblock Constructing high-dimensional neural network potentials: a tutorial
  review.
\newblock {\em International Journal of Quantum Chemistry}, 115(16):1032--1050,
  2015.

\bibitem{behler-fingerprints-2011}
J{\"o}rg Behler.
\newblock Atom-centered symmetry functions for constructing high-dimensional
  neural network potentials.
\newblock {\em The Journal of chemical physics}, 134(7), 2011.

\bibitem{si}
Suplementary Material.

\bibitem{Zhang-GL-2019}
Huaqing Zhang, Jian Wang, Zhanquan Sun, Jacek~M Zurada, and Nikhil~R Pal.
\newblock Feature selection for neural networks using group lasso
  regularization.
\newblock {\em IEEE Transactions on Knowledge and Data Engineering},
  32(4):659--673, 2019.

\bibitem{Feng-2019}
Jean Feng and Noah Simon.
\newblock Sparse-input neural networks for high-dimensional nonparametric
  regression and classification.
\newblock {\em arXiv preprint arXiv:1711.07592}, 2017.

\bibitem{N2P2}
Andreas Singraber, J{\"o}rg Behler, and Christoph Dellago.
\newblock Library-based lammps implementation of high-dimensional neural
  network potentials.
\newblock {\em Journal of chemical theory and computation}, 15(3):1827--1840,
  2019.

\bibitem{cur-implementation}
Alexander Goscinski, Guillaume Fraux, Giulio Imbalzano, and Michele Ceriotti.
\newblock The role of feature space in atomistic learning.
\newblock {\em Machine Learning: Science and Technology}, 2(2):025028, 2021.

\bibitem{ovito}
Alexander Stukowski.
\newblock Visualization and analysis of atomistic simulation data with
  ovito--the open visualization tool.
\newblock {\em Modelling and simulation in materials science and engineering},
  18(1):015012, 2009.

\bibitem{ADAMW}
Ilya Loshchilov and Frank Hutter.
\newblock Decoupled weight decay regularization.
\newblock {\em arXiv preprint arXiv:1711.05101}, 2017.

\bibitem{range-test}
Leslie~N Smith.
\newblock Cyclical learning rates for training neural networks.
\newblock In {\em 2017 IEEE winter conference on applications of computer
  vision (WACV)}, pages 464--472. IEEE, 2017.

\bibitem{LASSO-bias}
Jason~D Lee, Dennis~L Sun, Yuekai Sun, and Jonathan~E Taylor.
\newblock Exact post-selection inference, with application to the lasso.
\newblock {\em Annals of Statistics}, 44(3):907--927, 2016.

\bibitem{lr-schedule}
Ilya Loshchilov and Frank Hutter.
\newblock Sgdr: Stochastic gradient descent with warm restarts.
\newblock {\em arXiv preprint arXiv:1608.03983}, 2016.

\bibitem{B-atomic-structure}
Si-Da Huang, Cheng Shang, Pei-Lin Kang, and Zhi-Pan Liu.
\newblock Atomic structure of boron resolved using machine learning and global
  sampling.
\newblock {\em Chemical science}, 9(46):8644--8655, 2018.

\bibitem{stability-2023}
Xiang Fu, Zhenghao Wu, Wujie Wang, Tian Xie, Sinan Keten, Rafael
  Gomez-Bombarelli, and Tommi Jaakkola.
\newblock Forces are not enough: Benchmark and critical evaluation for machine
  learning force fields with molecular simulations.
\newblock {\em arXiv preprint arXiv:2210.07237}, 2022.

\end{thebibliography}

\renewcommand{\thefigure}{S\arabic{figure}}
\renewcommand{\thetable}{S\Roman{table}}

\begin{center}
	{\Large Supplementary Information File}
\end{center}
\section{Symmetry Function Parameters}
\label{Sec:LJ-SF-Params}
Tables \ref{tab:LJ-SF-params}, \ref{tab:Al-SF-params} and \ref{tab:B-SF-params} contain the symmetry function parameters used for the Lennard Jones system, Al, and B respectively.
Definitions of the symmetry functions, and their parameters, are found in the main text.

\begin{table}[h]
	\centering
	\caption{\label{tab:LJ-SF-params}Parameter values of symmetry functions used for the Lennard Jones system.}
	\begin{tabular}{ccccccc|ccccccc}\hline
		index & type & $\eta$ & $\Lambda$ & $\zeta$ & $r_s$ & $r_c$ & index & type & $\eta$ & $\Lambda$ & $\zeta$ & $r_s$ & $r_c$\\\hline
		$1$ & $G^2$ & $0.12755$ & $-$ & $-$ & $0.0$ & $2.8$ & $13$ & $G^5$ & $0.12755$ & $-1.0$ & $1.0$ & $0.0$ & $2.8$\\
		$2$ & $G^2$ & $0.24281$ & $-$ & $-$ & $0.0$ & $2.8$ & $14$ & $G^5$ & $0.12755$ & $1.0$ & $1.0$ & $0.0$ & $2.8$\\
		$3$ & $G^2$ & $0.46223$ & $-$ & $-$ & $0.0$ & $2.8$ & $15$ & $G^5$ & $0.12755$ & $-1.0$ & $2.0$ & $0.0$ & $2.8$\\
		$4$ & $G^2$ & $0.87993$ & $-$ & $-$ & $0.0$ & $2.8$ & $16$ & $G^5$ & $0.12755$ & $1.0$ & $2.0$ & $0.0$ & $2.8$\\
		$5$ & $G^2$ & $1.6751$ & $-$ & $-$ & $0.0$ & $2.8$ & $17$ & $G^5$ & $0.12755$ & $-1.0$ & $4.0$ & $0.0$ & $2.8$\\
		$6$ & $G^2$ & $3.1888$ & $-$ & $-$ & $0.0$ & $2.8$ & $18$ & $G^5$ & $0.12755$ & $1.0$ & $4.0$ & $0.0$ & $2.8$\\
		$7$ & $G^2$ & $3.858$ & $-$ & $-$ & $0.5$ & $2.8$ & $19$ & $G^5$ & $0.12755$ & $-1.0$ & $8.0$ & $0.0$ & $2.8$\\
		$8$ & $G^2$ & $3.858$ & $-$ & $-$ & $0.86$ & $2.8$ & $20$ & $G^5$ & $0.12755$ & $1.0$ & $8.0$ & $0.0$ & $2.8$\\
		$9$ & $G^2$ & $3.858$ & $-$ & $-$ & $1.22$ & $2.8$ & $21$ & $G^5$ & $0.12755$ & $-1.0$ & $16.0$ & $0.0$ & $2.8$\\
		$10$ & $G^2$ & $3.858$ & $-$ & $-$ & $1.58$ & $2.8$ & $22$ & $G^5$ & $0.12755$ & $1.0$ & $16.0$ & $0.0$ & $2.8$\\
		$11$ & $G^2$ & $3.858$ & $-$ & $-$ & $1.94$ & $2.8$\\
		$12$ & $G^2$ & $3.858$ & $-$ & $-$ & $2.3$ & $2.8$\\\hline
	\end{tabular}
\end{table}

\begin{table}
	\centering
	\caption{\label{tab:Al-SF-params}Parameter values of symmetry functions used for Al.}
	\begin{tabular}{ccccccc|ccccccc}\hline
		index & type & $\eta$ & $\Lambda$ & $\zeta$ & $r_s$ & $r_c$ & index & type & $\eta$ & $\Lambda$ & $\zeta$ & $r_s$ & $r_c$\\\hline
		$1$ & $G^2$ & $0.0013$ & $-$ & $-$ & $0.0$ & $6.8$ & $13$ & $G^5$ & $0.0013$ & $-1.0$ & $1.0$ & $0.0$ & $6.8$\\
		$2$ & $G^2$ & $0.04$ & $-$ & $-$ & $0.0$ & $6.8$ & $14$ & $G^5$ & $0.0013$ & $1.0$ & $1.0$ & $0.0$ & $6.8$\\
		$3$ & $G^2$ & $0.1$ & $-$ & $-$ & $0.0$ & $6.8$ & $15$ & $G^5$ & $0.0013$ & $-1.0$ & $2.0$ & $0.0$ & $6.8$\\
		$4$ & $G^2$ & $0.26$ & $-$ & $-$ & $0.0$ & $6.8$ & $16$ & $G^5$ & $0.0013$ & $1.0$ & $2.0$ & $0.0$ & $6.8$\\
		$5$ & $G^2$ & $0.66$ & $-$ & $-$ & $0.0$ & $6.8$ & $17$ & $G^5$ & $0.0013$ & $-1.0$ & $4.0$ & $0.0$ & $6.8$\\
		$6$ & $G^2$ & $1.85$ & $-$ & $-$ & $0.0$ & $6.8$ & $18$ & $G^5$ & $0.0013$ & $1.0$ & $4.0$ & $0.0$ & $6.8$\\
		$7$ & $G^2$ & $1.85$ & $-$ & $-$ & $1.017$ & $6.8$ & $19$ & $G^5$ & $0.0013$ & $-1.0$ & $16.0$ & $0.0$ & $6.8$\\
		$8$ & $G^2$ & $1.85$ & $-$ & $-$ & $2.035$ & $6.8$ & $20$ & $G^5$ & $0.0013$ & $1.0$ & $16.0$ & $0.0$ & $6.8$\\
		$9$ & $G^2$ & $1.85$ & $-$ & $-$ & $3.052$ & $6.8$ & $21$ & $G^5$ & $0.0013$ & $-1.0$ & $64.0$ & $0.0$ & $6.8$\\
		$10$ & $G^2$ & $1.85$ & $-$ & $-$ & $4.07$ & $6.8$ & $22$ & $G^5$ & $0.0013$ & $1.0$ & $64.0$ & $0.0$ & $6.8$\\
		$11$ & $G^2$ & $1.85$ & $-$ & $-$ & $5.087$ & $6.8$\\
		$12$ & $G^2$ & $1.85$ & $-$ & $-$ & $6.105$ & $6.8$\\\hline
	\end{tabular}
\end{table}

\begin{table}
	\centering
	\caption{\label{tab:B-SF-params}Parameter values of symmetry functions used for B.}
	\begin{tabular}{ccccccc|ccccccc}\hline
		index & type & $\eta$ & $\Lambda$ & $\zeta$ & $r_s$ & $r_c$ & index & type & $\eta$ & $\Lambda$ & $\zeta$ & $r_s$ & $r_c$\\\hline
		$1$ & $G^2$ & $0.012356$ & $-$ & $-$ & $0.0$ & $5.3$ & $31$ & $G^5$ & $0.12$ & $-1.0$ & $8.0$ & $0.0$ & $5.3$\\
		$2$ & $G^2$ & $0.033587$ & $-$ & $-$ & $0.0$ & $5.3$ & $32$ & $G^5$ & $0.12$ & $1.0$ & $8.0$ & $0.0$ & $5.3$\\
		$3$ & $G^2$ & $0.091298$ & $-$ & $-$ & $0.0$ & $5.3$ & $33$ & $G^5$ & $0.12$ & $-1.0$ & $16.0$ & $0.0$ & $5.3$\\
		$4$ & $G^2$ & $0.24817$ & $-$ & $-$ & $0.0$ & $5.3$ & $34$ & $G^5$ & $0.12$ & $1.0$ & $16.0$ & $0.0$ & $5.3$\\
		$5$ & $G^2$ & $0.67461$ & $-$ & $-$ & $0.0$ & $5.3$ & $35$ & $G^5$ & $0.12$ & $-1.0$ & $64.0$ & $0.0$ & $5.3$\\
		$6$ & $G^2$ & $1.8338$ & $-$ & $-$ & $0.0$ & $5.3$ & $36$ & $G^5$ & $0.12$ & $1.0$ & $64.0$ & $0.0$ & $5.3$\\
		$7$ & $G^2$ & $3.6675$ & $-$ & $-$ & $0.75714$ & $5.3$ & $37$ & $G^5$ & $0.5$ & $-1.0$ & $1.0$ & $0.0$ & $5.3$\\
		$8$ & $G^2$ & $3.6675$ & $-$ & $-$ & $1.5143$ & $5.3$ & $38$ & $G^5$ & $0.5$ & $1.0$ & $1.0$ & $0.0$ & $5.3$\\
		$9$ & $G^2$ & $3.6675$ & $-$ & $-$ & $2.2714$ & $5.3$ & $39$ & $G^5$ & $0.5$ & $-1.0$ & $2.0$ & $0.0$ & $5.3$\\
		$10$ & $G^2$ & $3.6675$ & $-$ & $-$ & $3.0286$ & $5.3$ & $40$ & $G^5$ & $0.5$ & $1.0$ & $2.0$ & $0.0$ & $5.3$\\
		$11$ & $G^2$ & $3.6675$ & $-$ & $-$ & $3.7857$ & $5.3$ & $41$ & $G^5$ & $0.5$ & $-1.0$ & $4.0$ & $0.0$ & $5.3$\\
		$12$ & $G^2$ & $3.6675$ & $-$ & $-$ & $4.5429$ & $5.3$ & $42$ & $G^5$ & $0.5$ & $1.0$ & $4.0$ & $0.0$ & $5.3$\\
		$13$ & $G^5$ & $0.01$ & $-1.0$ & $1.0$ & $0.0$ & $5.3$ & $43$ & $G^5$ & $0.5$ & $-1.0$ & $8.0$ & $0.0$ & $5.3$\\
		$14$ & $G^5$ & $0.01$ & $1.0$ & $1.0$ & $0.0$ & $5.3$ & $44$ & $G^5$ & $0.5$ & $1.0$ & $8.0$ & $0.0$ & $5.3$\\
		$15$ & $G^5$ & $0.01$ & $-1.0$ & $2.0$ & $0.0$ & $5.3$ & $45$ & $G^5$ & $0.5$ & $-1.0$ & $16.0$ & $0.0$ & $5.3$\\
		$16$ & $G^5$ & $0.01$ & $1.0$ & $2.0$ & $0.0$ & $5.3$ & $46$ & $G^5$ & $0.5$ & $1.0$ & $16.0$ & $0.0$ & $5.3$\\
		$17$ & $G^5$ & $0.01$ & $-1.0$ & $4.0$ & $0.0$ & $5.3$ & $47$ & $G^5$ & $0.5$ & $-1.0$ & $64.0$ & $0.0$ & $5.3$\\
		$18$ & $G^5$ & $0.01$ & $1.0$ & $4.0$ & $0.0$ & $5.3$ & $48$ & $G^5$ & $0.5$ & $1.0$ & $64.0$ & $0.0$ & $5.3$\\
		$19$ & $G^5$ & $0.01$ & $-1.0$ & $8.0$ & $0.0$ & $5.3$ & $49$ & $G^5$ & $0.5$ & $-1.0$ & $1.0$ & $4.85$ & $5.3$\\
		$20$ & $G^5$ & $0.01$ & $1.0$ & $8.0$ & $0.0$ & $5.3$ & $50$ & $G^5$ & $0.5$ & $1.0$ & $1.0$ & $4.85$ & $5.3$\\
		$21$ & $G^5$ & $0.01$ & $-1.0$ & $16.0$ & $0.0$ & $5.3$ & $51$ & $G^5$ & $0.5$ & $-1.0$ & $2.0$ & $4.85$ & $5.3$\\
		$22$ & $G^5$ & $0.01$ & $1.0$ & $16.0$ & $0.0$ & $5.3$ & $52$ & $G^5$ & $0.5$ & $1.0$ & $2.0$ & $4.85$ & $5.3$\\
		$23$ & $G^5$ & $0.01$ & $-1.0$ & $64.0$ & $0.0$ & $5.3$ & $53$ & $G^5$ & $0.5$ & $-1.0$ & $4.0$ & $4.85$ & $5.3$\\
		$24$ & $G^5$ & $0.01$ & $1.0$ & $64.0$ & $0.0$ & $5.3$ & $54$ & $G^5$ & $0.5$ & $1.0$ & $4.0$ & $4.85$ & $5.3$\\
		$25$ & $G^5$ & $0.12$ & $-1.0$ & $1.0$ & $0.0$ & $5.3$ & $55$ & $G^5$ & $0.5$ & $-1.0$ & $8.0$ & $4.85$ & $5.3$\\
		$26$ & $G^5$ & $0.12$ & $1.0$ & $1.0$ & $0.0$ & $5.3$ & $56$ & $G^5$ & $0.5$ & $1.0$ & $8.0$ & $4.85$ & $5.3$\\
		$27$ & $G^5$ & $0.12$ & $-1.0$ & $2.0$ & $0.0$ & $5.3$ & $57$ & $G^5$ & $0.5$ & $-1.0$ & $16.0$ & $4.85$ & $5.3$\\
		$28$ & $G^5$ & $0.12$ & $1.0$ & $2.0$ & $0.0$ & $5.3$ & $58$ & $G^5$ & $0.5$ & $1.0$ & $16.0$ & $4.85$ & $5.3$\\
		$29$ & $G^5$ & $0.12$ & $-1.0$ & $4.0$ & $0.0$ & $5.3$ & $59$ & $G^5$ & $0.5$ & $-1.0$ & $64.0$ & $4.85$ & $5.3$\\
		$30$ & $G^5$ & $0.12$ & $1.0$ & $4.0$ & $0.0$ & $5.3$ & $60$ & $G^5$ & $0.5$ & $1.0$ & $64.0$ & $4.85$ & $5.3$\\\hline
	\end{tabular}
\end{table}

\section{Test Errors}
\label{Sec:test-errors}
Tables \ref{tab:LJ-test}, \ref{tab:Al-test} and \ref{tab:B-test} show the average test errors of the models trained for each featureset discussed in the main text.

\begin{table}
	\centering
	\caption{\label{tab:LJ-test}
		Total number of features $(N)$, number of angular features $(N_{G^5})$, test error averaged over four models, and computational performance of selected features for LJ, for the original set of features and for the different feature selection methods discussed in the text.}
	\begin{tabular}{ccccc}\hline
		Method & $N$ & $N_{G^5}$ & RMSE ($10^{-3}\epsilon$/atom) & Benchmark (timesteps/s)\\\hline
		- & $22$ & $10$ & $9.73\pm0.08$ & $1.053$\\
		AGL & $7$ & $1$ & $8.59 \pm 0.08$ & $2.892$\\
		PC & $7$ & $1$ & $8.79\pm0.07$ & $2.885$\\
		CUR & $7$ & $2$ & $23.6\pm0.1$ & -\\\hline
	\end{tabular}
\end{table}

\begin{table}
	\centering
	\caption{\label{tab:Al-test}
		Total number of features $(N)$, number of angular features $(N_{G^5})$, test error averaged over four models, and computational performance of selected features for Al, for the original set of features and for the different feature selection methods with 10 respectively 7 chosen features. The last line shows the results using the feature set chosen by AGL for the LJ training.}
	\begin{tabular}{ccccc}\hline
		method & $N$ & $N_{G^5}$ & RMSE (meV/atom) & Benchmark (timesteps/s)\\\hline
		- & $22$ & $10$ & $2.07\pm0.17$ & $1.626$ \\
		AGL & $10$ & $2$ & $2.29\pm0.16$ & $3.582$ \\
		PC & $10$ & $1$ & $2.43\pm0.06$ & $4.183$ \\
		CUR & $10$ & $3$ & $2.44\pm0.35$ & $3.154$ \\\hline
		AGL & $7$ & $1$ & $2.78\pm0.07$ & $4.132$ \\
		PC & $7$ & $1$ & $3.06\pm0.27$ & $4.372$ \\
		CUR & $7$ & $2$ & $6.38\pm0.10$ & $2.900$ \\\hline
		AGL (LJ) & $7$ & $1$ & $2.67 \pm 0.05$ & $4.399$\\\hline
	\end{tabular}
\end{table}

\begin{table}
	\centering
	\caption{\label{tab:B-test}
		Total number of features $(N)$, number of angular features $(N_{G^5})$, test error averaged over four models, and computational performance of selected features for B, for different feature selection methods and original set of features.}.
	\begin{tabular}{ccccc}\hline
		method & $N$ & $N_{G^5}$ & RMSE (meV/atom) & Benchmark (timesteps/s)\\\hline
		- & $60$ & $48$ & $6.95\pm0.09$ & $0.315$\\
		AGL & $16$ & $10$ & $8.12\pm0.03$ & $1.13$\\
		PC & $16$ & $8$ & $10.8\pm0.12$ & -\\
		CUR & $16$ & $11$ & $10.4\pm0.08$ & -\\\hline
	\end{tabular}
\end{table}
\end{document}